# Prediction of the
# Infrared Absorbance Intensities and Frequencies of Hydrocarbons:
# A Message Passing Neural Network Approach

**Maliheh Shaban Tameh, Veaceslav Coropceanu,**

**Thomas A.R. Purcell[*], and Jean-Luc Brédas[*]**

Department of Chemistry and Biochemistry

The University of Arizona

Tucson, Arizona 85721-0041

* Emails: purcellt@arizona.edu; jlbredas@arizona.edu

## Abstract

Accurately and efficiently predicting the infrared (IR) spectra of a molecule can provide insights into the structure-properties relationships of molecular species, which has led to a proliferation of machine learning tools designed for this purpose. However, earlier studies have focused primarily on obtaining normalized IR spectra, which limits their potential for a comprehensive analysis of molecular behavior in the IR range. For instance, to fully understand and predict the optical properties, such as the transparency characteristics, it is necessary to predict the molar absorptivity IR spectra instead. Here, we propose a graph-based communicative message passing neural network (CMPNN) algorithm that can predict both the peak positions and absolute intensities corresponding to density functional theory (DFT) calculated molar absorptivities in the IR domain. By modifying existing spectral loss functions, we show that our method is able to predict with DFT-accuracy level the IR molar absorptivities of a series of hydrocarbons containing up to ten carbon atoms and apply the model to a set of larger molecules. We also compare the predicted spectra with those generated by the direct message passing neural network (DMPNN). The results suggest that both algorithms demonstrate similar predictive capabilities for hydrocarbons, indicating that either model could be effectively used in future research on spectral prediction for such systems.

# Introduction

Infrared (IR) spectroscopy has long been used to identify organic moieties and to determine structure-properties relationships by observing the emergence, disappearance, or shift of characteristic signals throughout a chemical process. The information based on IR spectroscopy can also be utilized to design new materials. For instance, identifying molecules with no or weakly IR active vibrational modes in the $800 - 1250 \text{ cm}^{-1}$ energy range is paramount for the design of new systems for long-wave IR (LWIR) imaging applications.[1-3] A targeted search for hydrocarbons that do not absorb significantly in this range can be significantly enhanced using predictive models. These models should not only forecast LWIR spectra but also estimate LWIR transmittance values across a vast chemical space, offering a robust scientific approach for identifying potential candidates.

Because of their ease of calculation and exact structural matching, simulated IR spectra have been increasingly used to complement their experimentally measured counterparts. In particular, density functional theory (DFT) has emerged as a popular tool for calculating the gas-phase IR spectra of molecules due to a reasonable balance between accuracy and computational cost[4]; it was recently used as a screening tool to simulate the IR spectra of polymeric transmissive materials.[1, 3] Despite the accuracy, efficiency, and reliability of DFT in simulating the vibrational spectra of molecular systems, the overall cost of screening millions of potential molecules remains prohibitive. Machine learning (ML) models that can predict the gas-phase IR spectra of molecules are a way to solve this issue.

Recently, ML algorithms have demonstrated strong potential in describing the molecular properties in the context of chemical,[5] drug design,[6] or materials science problems.[7] In particular, multiple ML algorithms can now predict IR spectra.[8-14] However, in these earlier applications, the training data have focused on the relative intensities of *normalized* spectra, which prevents these methodologies from predicting the IR transparency of the molecules. Here, we demonstrate the application of deep neural network (DNN) algorithms for the prediction of not only the frequencies but importantly also the absolute intensities of the IR absorbance spectra of hydrocarbons.

Neural Network (NN) architectures have achieved high performance in the prediction of normalized IR spectra for various types of datasets.[15-17] Among them, message passing neural network (MPNN) algorithms have been promising for both IR and ultraviolet-visible spectra.[17, 18] McGill *et al.*[19] predicted IR spectra using the directed MPNN (DMPNN)[20, 21] architecture introduced by Yang *et al.* through the Chemprop-IR package.[19] In contrast to generic MPNN, a DMPNN architecture centers on the bonds instead of the atoms in a molecular graph, thereby avoiding unnecessary loops in the message passing trajectory. Although this development enables a relatively inexpensive and accurate prediction of IR spectra, their focus on replicating experimental spectra from multiple sources makes it impossible to model exact absorbance values.[22]

In this work, we extend the current state-of-the-art MPNN protocol to predict the absolute intensities of IR spectra, which is an essential step in assessing transmittance efficiency in specific IR windows. To do so, we calculate the spectra using DFT with the B3LYP[23] functional and use the data to train both a DMPNN and a communicative message passing neural network (CMPNN).

CMPNN is an architecture proposed by Song *et al.*[24] that introduces a node-edge interaction procedure and strengthens the message interaction among atoms and bonds, which has been shown to outperform DMPNN for property-prediction tasks.[25-28] Using the calculated spectra instead of the experimental ones provides us with a controllable way of introducing absolute intensities.[4, 29, 30] Also, we propose a modification to the spectral information divergence (SID) loss function to account for non-normalized data. We test the models on a set of larger molecules to compare the performance of CMPNN and DMPNN for IR-spectra predictions. Additionally, we investigate the impact of specific structural motifs on the performance and generalizability of the DMPNN model; by performing a principal component analysis, we are able to identify the most significant features for spectral prediction.

## Methodology

*Database.* We select our database, using the SMILES[31] format as the identifier, as a subclass of the generated databases (GDBs)[32, 33] by Reymond and co-workers. GDB-11 collects organic molecules with up to 11 atoms of C, N, O, or F (with rules expressing chemical stability and synthetic feasibility). Our dataset includes 27,845 hydrocarbons with ten carbon atoms; 5,714 hydrocarbons with nine carbon atoms; 1,274 hydrocarbons with eight carbon atoms; 290 hydrocarbons with seven carbon atoms; and 115 hydrocarbons with two to six carbon atoms. All the selected molecules are unsaturated, include no other heavy atom than carbon, are neutral and closed-shell, and can incorporate bicyclic (bridged, fused), tricyclic, or polycyclic structures.

*Automation of DFT data generation.* To facilitate the building of a large-scale dataset for training machine learning models, we create an automated data generation process as recently described in our previous work.[3] The process comprises a high-throughput computational workflow, implemented in python, with the following steps, summarized in Figure 1: (1) generate an initial geometry from the SMILES representations of the molecules, (2) pre-optimize the geometry using the universal force field[34] (UFF) implemented in RDKit,[35] (3) optimize the geometry and calculate its vibrational modes with Gaussian16[36] using the B3LYP[23, 37, 38] functional and the 6-31g(d,p) basis set, and (4) generate the IR molar absorptivity spectra. We note that the calculated IR spectra represent only the harmonic vibrational frequencies with no anharmonic interactions included. The computed stick-like spectra are broadened with a Lorentzian band shape with a $\gamma$ (half-width at half-height) value of 10 cm$^{-1}$, for which the molar absorption coefficient is given as: [39]

$$\varepsilon(v) = \frac{1}{1000 \ln(10)\pi} \sum_i \frac{v}{v_i} \frac{I_i \gamma}{(v - v_i)^2 + \gamma^2}$$

Here, the $i$ subscript refers to a given normal mode; $v_i$ (cm$^{-1}$) and $I_i$ (cm mol$^{-1}$) are the frequency and IR intensity of the normal mode, respectively. We store the molar absorbance spectra with 1801 frequency points evenly spaced by 2 cm$^{-1}$ between 400 and 4,000 cm$^{-1}$.

Out of the 35,238 molecules considered initially, 31,570 successfully completed this workflow and are considered below (we emphasize that all these molecules do not exhibit any imaginary frequencies). The associated optimized geometries, molecular frequencies, absorption intensities, and IR molar absorptivity spectra of each molecule are used in our subsequent investigations.

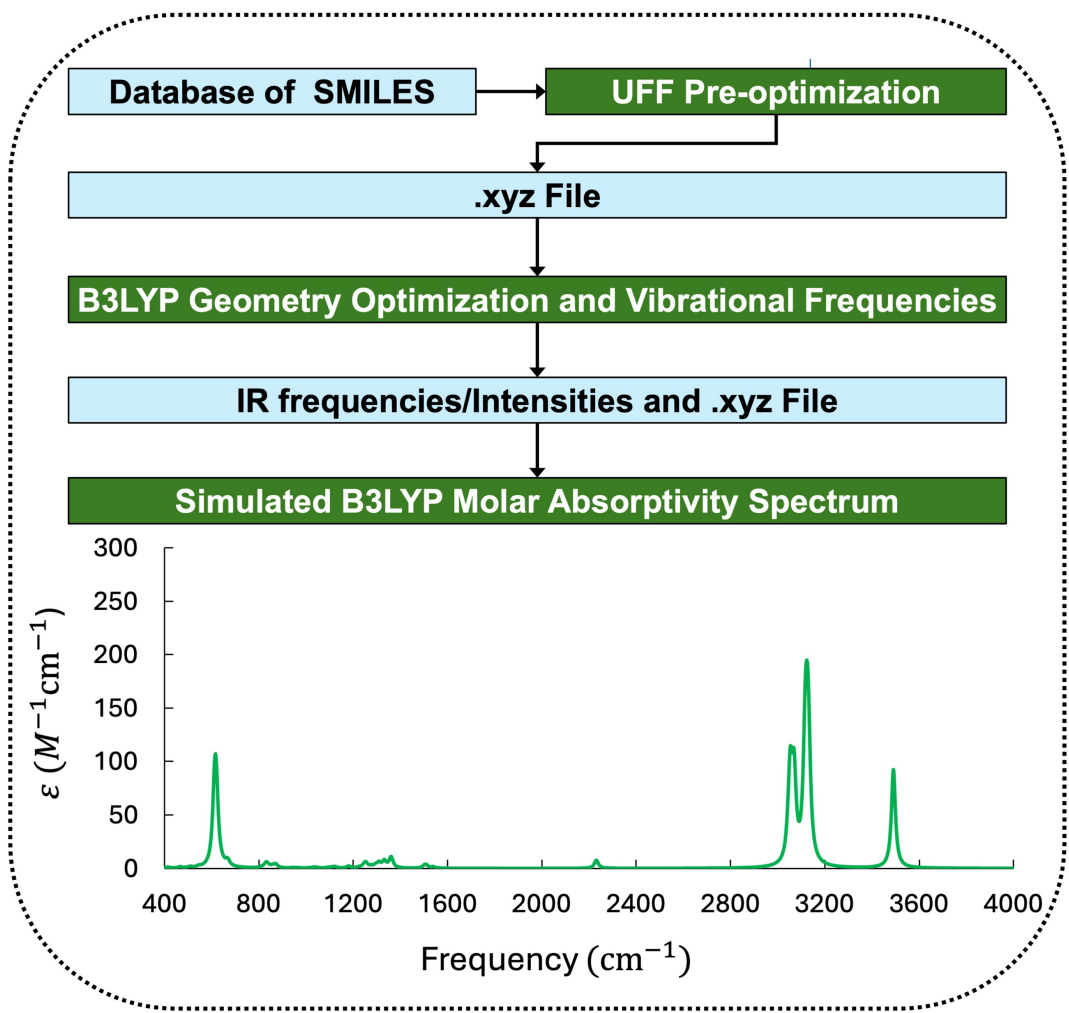

**Figure 1**. Computational workflow to generate the dataset, from molecule enumeration, pre-optimization, to high-throughput DFT calculations to obtain the optimized molecules, the IR vibrational frequencies and IR intensities, and the molar absorptivity spectra.

*Message passing architecture.* A MPNN framework operates on directed graphs $G(V, E)$, where $V$ represents the set of nodes (atoms) and $E$ represents the set of edges (bonds). Each node $v, w, u \in V$ corresponds to an atom in the graph, and an edge $\boldsymbol{e}_{v,w} \in E$ denotes a bond between atom $v$ and atom $w$. The set $N(v)$ refers to the neighboring nodes of node $v$.

In this work, we integrate the existing CMPNN,[24] an MPNN that includes node-edge interaction approach, into the open-source Chemprop-IR software[19] to evaluate whether the improved property description reported from this framework[24] also holds for spectral predictions. CMPNN has two phases, a message passing phase and a readout phase. In the former, the node and edge information of the molecule is propagated at each iteration for T time steps to build a neural representation of the molecule. In the readout phase, a readout function is used to make a prediction based on the final hidden states by computing a feature vector for the whole graph. This representation learning approach is a directed message passing neural network that treats the molecular graph as an edge-oriented direct structure, enhanced by the node-edge interaction module. In this module, both the edge and node embeddings are updated interactively during the molecular embedding generation process. This means that both the representations of individual nodes and edges are updated based on the information received from the neighboring nodes; thus, the model can use both atomic and bond information when making predictions or performing downstream tasks on the graph. This interaction leads to more powerful models capable of capturing complex relationships and patterns in the graph data. Notably, the CMPNN operates on node-hidden states $\boldsymbol{h}(v)$, edge-hidden states $\boldsymbol{h}(\boldsymbol{e}_{v,w})$, and messages $\boldsymbol{m}(\boldsymbol{e}_{v,w})$ and $\boldsymbol{m}(v)$,[24] as opposed to the DMPNN,[20] which relies on edge-hidden states $\boldsymbol{h}(\boldsymbol{e}_{v,w})$ and messages $\boldsymbol{m}(\boldsymbol{e}_{v,w})$, and the MPNN[40], which focuses on node-hidden states $\boldsymbol{h}(v)$ and messages $\boldsymbol{m}(v)$.

Chemprop-IR predicts the IR spectra from a given graph as input in an end-to-end fashion using a DMPNN.[19, 40] Chemprop-IR was initially trained on *normalized* spectra to learn the experimental IR absorption spectra, which are normally reported with arbitrary units. While this approach is necessary when incorporating data from non-standardized sources, it makes impractical predicting a variety of physical and chemical properties, such as optical transmittance. In our protocol, we use the input (target) IR molar absorptivity spectra in the form of unnormalized absorbance quantities calculated from DFT.

In this work, the CMPNN model architecture consists of a directed MPNN and a feedforward neural network (FFNN) readout. The framework uses RDKit to generate molecular graphs from the SMILES representation of the molecules,[41] extracting features that are then used to predict the IR molar absorptivity spectra. These features form hidden states, which are propagated through the network during each message-passing iteration. A combination of atom and bond features is introduced during the message-passing steps to generate the learned molecular graph embedding. This encoding of molecules with CMPNN occurs prior to the FFNN readout. In the directed message passing stage, the messages and hidden states are associated with directed edges and node-edge interactions. We summarize the workflow in Figure 2, which consists of 5 modules: (1) extraction of bond and atom features, (2) a node-edge communicative message passing neural network operating on the directed graphs to learn atomic graph embeddings, where messages from incoming bonds and the current atom's information are gathered to update the atom's representation at step $t + 1$, (3) aggregation of these atomic graph embeddings into molecular graph embeddings, (4) concatenation of molecular graph embeddings from different molecules, followed by the application of a FFNN network to predict IR molar absorbance intensity vectors. The protocol is valid for each $N$ molecular graph in the dataset. The number of message passing steps is called depth and the size of bond message vectors is called hidden size. Further details on the directed message passing architecture can be found in the works of Yang *et al.*[20] and Song *et al.*[24], with a summary provided in the SI.

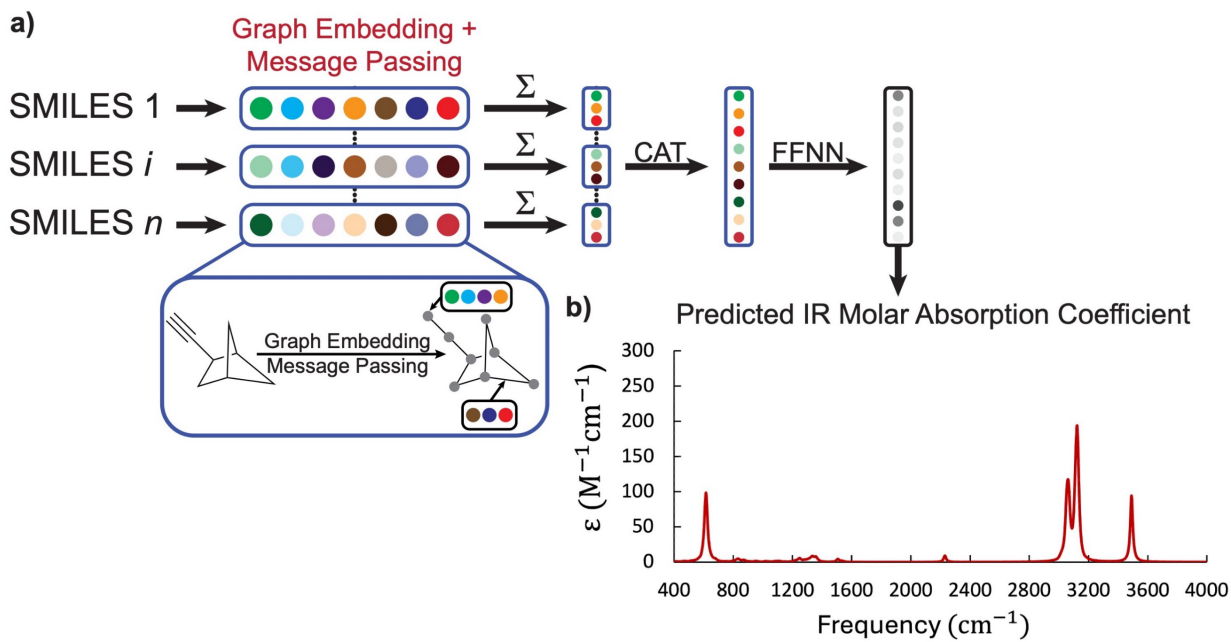

**Figure 2**. a) Schematic representation of the neural network predictions for the IR molar absorption coefficients. Each SMILES, $i$, of the training set of $n$ molecules is converted into molecular graphs and the individual vertexes and edges represent the atoms and bonds of the molecule, respectively. Features are then added to the graph and updated using a message passing formalism to capture the chemical environment, and aggregated to get the final molecular graph embedding, shown here by the colored circles. All embeddings are concatenated (CAT) and passed to a Feedfoward Neural Network (FFNN) that predicts the molar absorption coefficient for the molecules in the specified IR range represented by the black and white dots. b) The ML-predicted molar absorptivity spectrum is plotted using the predicted IR molar absorption coefficients over a range of frequencies. The plotted IR molar absorptivity spectrum corresponds to the example molecule in part (a).

The initial node features are the atomic features for each site and the initial edge features are the concatenation of the atomic features for an atom and the bond features of the bond between that atom and its neighboring atoms. The atomic features we consider are atomic number, atomic mass, the number of bonds associated with that atom, formal charge, the chirality, number of bonded hydrogen atoms, the hybridization, and whether the atom belongs to an aromatic ring. The bond features are the bond type, whether the bond is conjugated, whether the bond is involved in a ring, and the stereochemistry of the bond. All the atom and bond features are one-hot encodings except for atomic mass. A detailed description of the atomic and bond features is given in Table 1 and Table 2.

**Table 1**. Atom features as represented from Ref.[20]

| Feature | Description | size |
|---|---|---|
| Atom type | Type of atom (ex. C, N, O), by atomic number | 100 |
| # bonds | Number of bonds the atom is involved in | 6 |
| Formal charge | Integer electronic charge assigned to atom | 5 |
| Chirality | Unspecified, tetrahedral CW/CCW, or other | 4 |
| # Hs | Number of bonded hydrogen atoms | 5 |
| Hybridization | sp, sp2, sp3, sp3d, or $sp^3d^2$ | 5 |
| Aromaticity | Whether this atom is part of an aromatic system | 1 |
| Atomic mass | Mass of the atom, divided by 100 | 1 |

**Table 2**. Bond features as represented from Ref.[20]

| Feature | Description | size |
|---|---|---|
| Bond type | Single, double, triple, or aromatic | 4 |
| Conjugated | Whether the bond is conjugated | 1 |
| In ring | Whether the bond is part of a ring | 1 |
| Stereo | None, any, E/Z, or cis/trans | 6 |

***Loss function.*** We consider a modified spectral information divergence (SID) developed by Chang[42] as the loss function for both hyperparameter optimization and the metric for validation. The SID was demonstrated to be a good loss function for the Chemprop-IR protocol by McGill and coworkers.[19] Here, we normalize the SID value by the sum of the elements in the target spectra, $\Sigma_i(y_{target,i})$, and denote the modified metric as $\widehat{SID}$. This modification reverts to the original SID definition if used on normalized spectra and is defined as:

$$\widehat{SID}(\mathbf{y}_{pred}, \mathbf{y}_{target}) = \frac{1}{\Sigma_i(y_{target,i})} \Sigma_i \left( y_{pred,i} \ln \frac{y_{pred,i}}{y_{target,i}} + y_{target,i} \ln \frac{y_{target,i}}{y_{pred,i}} \right).$$

The $\mathbf{y}_{pred}$ and $\mathbf{y}_{target}$ are vectors of the non-normalized predicted and target spectra with positive absorbance values. We use $\widehat{SID}$ for all models generated in this work to ensure that all loss estimates are not biased by the integrated magnitude of the molar absorptivity spectra. The positivity of the predicted spectra is imposed by an exponential activation function to the FFNN output.

***Training protocol.*** We use the generated DFT IR molar absorptivity spectra with 1,801 elements of the molar absorption coefficient at 2 cm$^{-1}$ intervals between $400 - 4000$ cm$^{-1}$ as the target spectra to train the models. We split the dataset into a training set (80% of the data), a validation set (10% of the data), and test set (10% of the data). We then train an ensemble of 10 independent submodels, each with a different set of initialized weight. The optimized hyperparameters are optimized against the validation set. We then average the predicted spectra from all submodels to get a final prediction of the spectrum for each molecule and evaluate the performance of the model on test set. This ensemble approach has been demonstrated to give reliable results.[19, 21, 43] The hyperparameters were optimized using Bayesian optimization, as implemented in the Hyperopt library.[44] Through the hyperparameter tuning, we optimize the number of message passing steps (depth) and the associated hidden size, FFN network hidden size and layers, and the batch size. The optimization is performed for 50 iterations for a single random data split. The hyperparameter tuning is performed on a single model, with no ensembling considered during the optimization process. We also manually tune the dropout ratio to a value of 0.001 to reduce overfitting. As for the learning rates, a maximum value of $7\times10^{-5}$ is used in the learning rate scheduler, and the final value is set as a ratio of the maximum learning rate. All other parameters are left as their default values.

The model operates with a depth of five and a hidden size with 2,000 elements during message passing step and 2 hidden layers each of size 2,000 in the FFN network. The number of training epochs is 250 and Adam is used as the optimizer.[45] The exponential activation enforcing positivity is applied. Using the optimized hyperparameters and training the entire ensemble of models, we achieve an ensemble validation $\widehat{SID}$ of 0.09.

*Metric for the prediction and evaluation of the spectra.* To get a guide to understand how well a predicted spectrum corresponds to the B3LYP-reference spectrum, we use the spectral information similarity (SIS) metric proposed by Green et al.[19] The SIS score is defined as:

$$SIS(\hat{y}_{pred}, \hat{y}_{target}) = \frac{1}{1 + \widehat{SID}(\hat{y}_{pred}, \hat{y}_{target})},$$

where $\hat{y}$ is the given spectrum convoluted with a Gaussian, $G_a$, with a width of $10 \text{ cm}^{-1}$

$$\hat{y} = G_a * y.$$

The width of $G_a$ is chosen to be consistent with Chemprop-IR and is smaller than the observed peak shifts. The Gaussian convolution provides a continuous representation of the deviation for peak shifts or distortion in the predicted and target spectra. A smaller width would have a lower SIS. A SIS score of 1 corresponds to a perfect match between the two spectra and a score of 0 corresponds to an infinite deviation between the two spectra.

***Comparison with DMPNN in predicting IR absorbance intensities.*** As a baseline comparison, we train a DMPNN using the exact same protocol as Chemprop-IR using our dataset and normalized-SID metric. We use the same training procedure for both networks to ensure compatibility of the results. The hyperparameters for the model are the same as the one trained on the CMPNN.

## Results and discussion

We now discuss the results of the IR intensity prediction by the two trained models using CMPNN and DMPNN in the Chemprop-IR framework. The two SIS scores, $SIS_{CMPNN}$ and $SIS_{DMPNN}$, are used to assess the prediction quality.

***(a) Analysis of the model on the test split.*** The performance distribution of the two models based on the CMPNN and DMPNN architectures in the test set is shown in Figure 3. An average SIS value of 0.950 [0.946] and a standard deviation of 0.039 [0.042] in the CMPNN [DMPNN] procedure indicate an outstanding prediction performance for both networks.

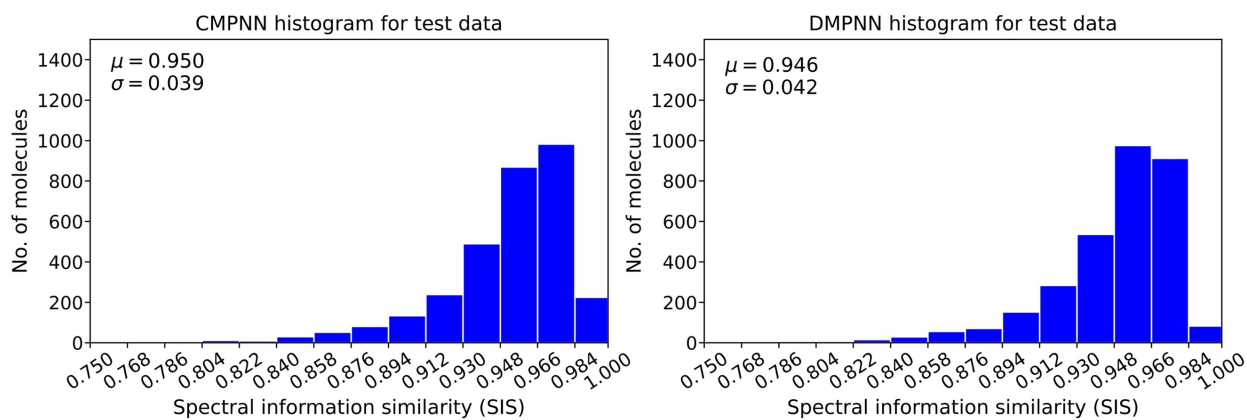

**Figure 3**. Distribution of the SIS score for the IR spectra prediction in the test split using CMPNN and DMPNN architectures; µ and σ represent the average value and standard deviation, respectively.

The quality of spectral prediction incorporates intensities, peak shapes, peak positions, continuous minor peaks, and distinctive major peaks. In the test split, 66.14 (62.84) percent of the CMPNN [DMPNN] predictions have SIS values greater than 0.948, which indicates a very accurate reproduction of the DFT B3LYP spectra,[45] as can be seen in Figure 4 a-c, with only minor disagreements between the computed and predicted spectra.

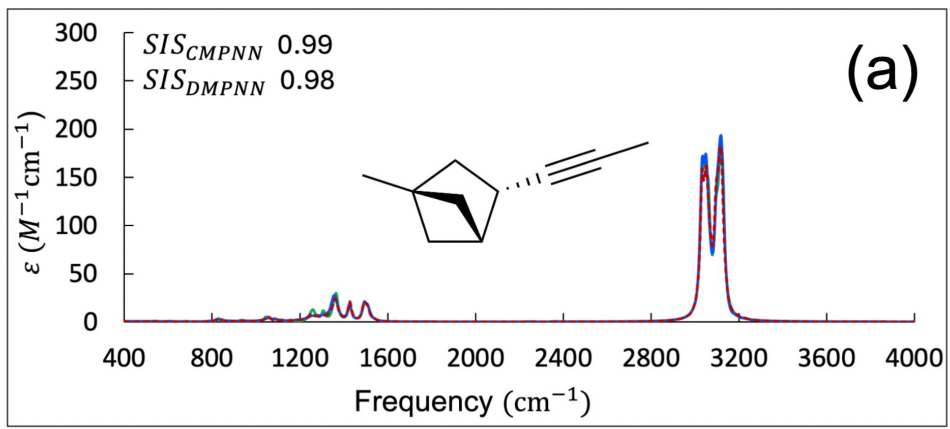

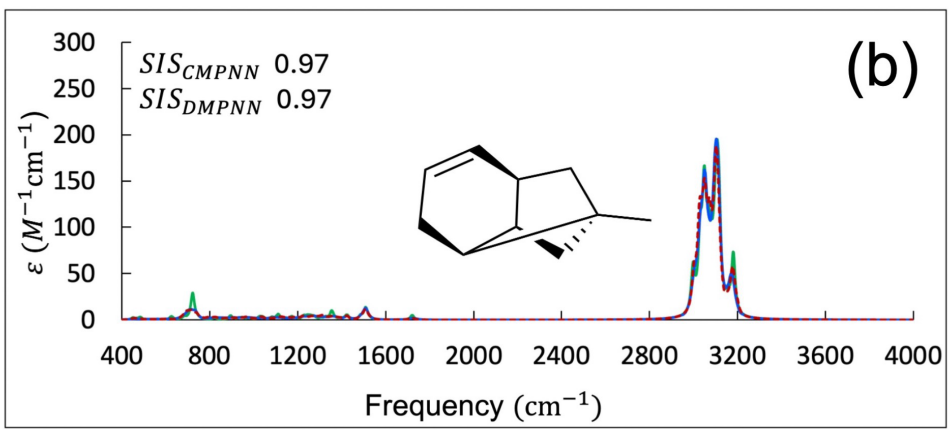

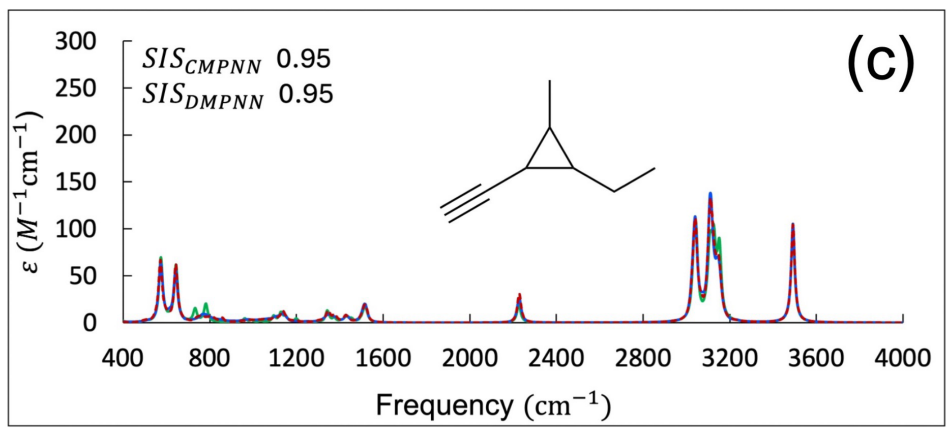

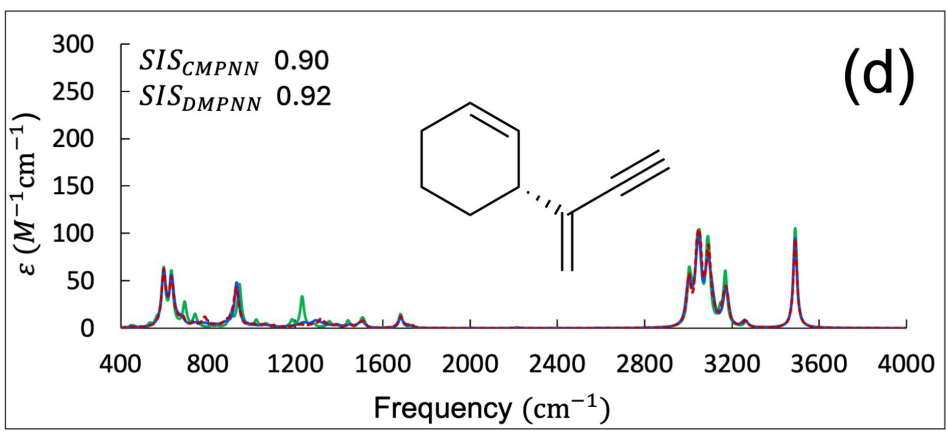

**Figure 4**. Comparison between the predicted CMPNN (red dashed) and DMPNN (blue solid) and the calculated B3LYP (green solid) IR molar absorptivity spectra in the 400 − 4000 cm$^{-1}$ window in the test set. The SIS values indicate the comparison between the B3LYP and related predicted spectra. The molecules for each spectrum are shown as insets in each plot.

Figure 4 presents a comparison of the predicted IR molar absorptivity spectra with their corresponding B3LYP results for some representative examples. Figure 4a and 4b illustrate accurate spectral predictions by both ML models (DMPNN and CMPNN), with SIS values above the average threshold of 0.95, indicating strong agreement with the DFT reference spectra. For these molecules, while there are minor differences with respect to the intensity of some peaks, the peak positions are preserved. Figure 4c demonstrates that a SIS value of 0.95 accurately reproduces all major peak clusters, while at least qualitatively describing the minor peaks. At a SIS of 0.9, differences in the intensities of the major peaks are expected, with the model also not being guaranteed to reproduce all minor peaks, as showcased for the ~1300 cm$^{-1}$ peak in Figure 4d.

For both ML models, about 90 percent of the data have SIS values above 0.90, below which moderate to significant peak discrepancies are observed, as illustrated in Figure 5. In particular, both ML models have difficulty in predicting the spectra of highly planar molecules, as demonstrated in Figure 5a and 5b. The ML models fail to reproduce the CH stretch portion of the spectra above 2800 cm$^{-1}$ and incorrectly predict the relative magnitudes of the other peaks. Additionally, some slight shifts in peak positions compared to the corresponding DFT results are observed, likely due to the higher symmetry of these 2D planar structures and the presence of delocalized π-electron systems. This is further illustrated in Figure 5c for benzene, where the CH stretches are red-shifted and the in-plane stretching modes at 1000 and 1500 cm$^{-1}$ do not appear. These findings highlight the challenges of modeling delocalized π systems, suggesting that an enhanced representation of such molecular features and a more targeted data augmentation, especially for planar structures, are likely to be required to enhance the accuracy of prediction in such instances.

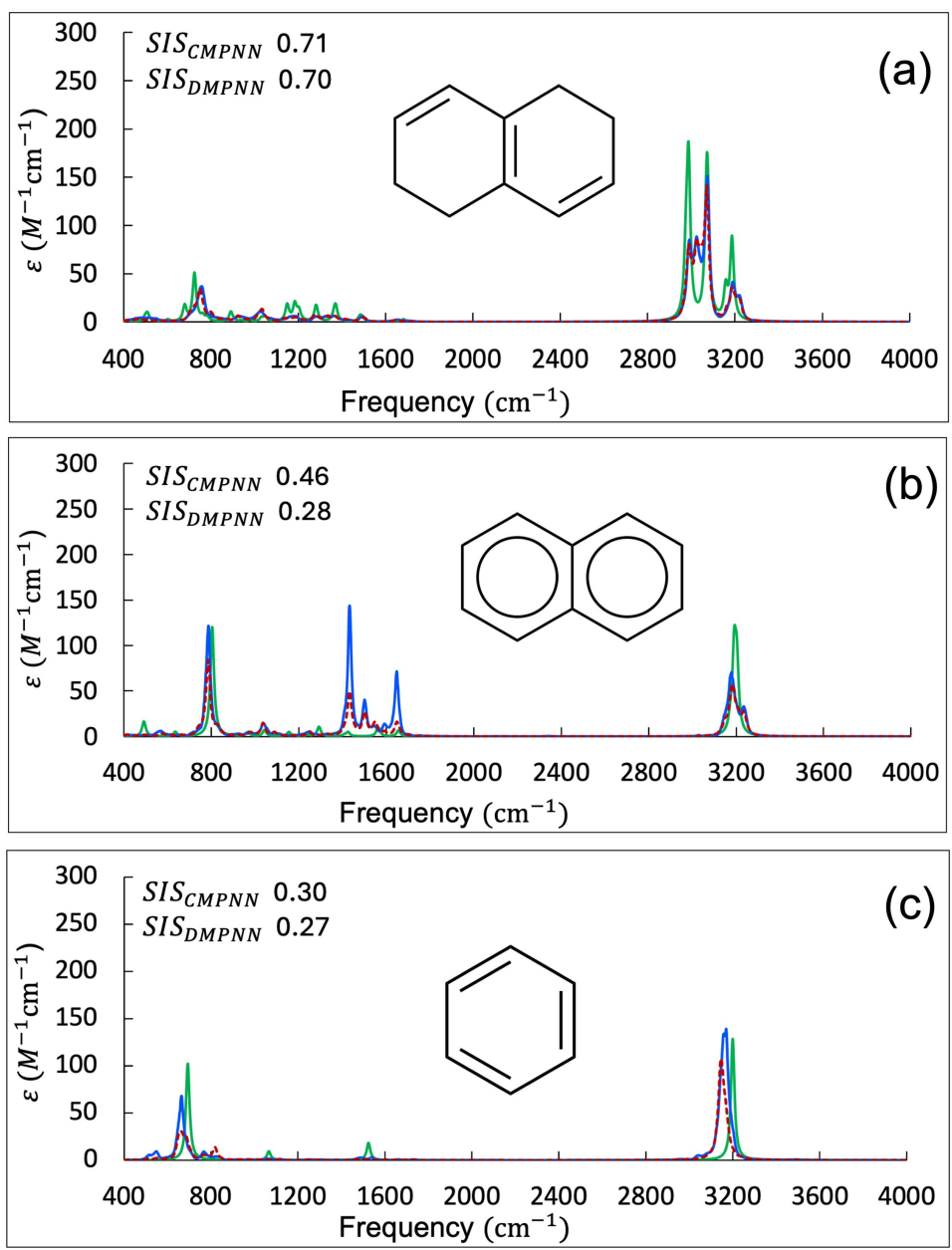

**Figure 5**. Comparison between the predicted CMPNN (red dashed) and DMPNN (blue solid) and the calculated B3LYP (green solid) IR molar absorptivity spectra in the 400 −

4000 cm$^{-1}$ window in the test set. The SIS values indicate the comparison between the B3LYP and related predicted spectra.

Excluding the weak performance for highly planar molecules, the comparison of the predicted spectra by CMPNN and DMPNN in Figure 4 highlights that both models produce very similar and accurate predictions of IR intensities. This finding suggests that strengthening the message interactions between edges and nodes in CMPNN does not necessarily improve the prediction of spectra within the MPNN architecture, for most hydrocarbon molecules. However, CMPNN does appear to outperform DMPNN when predicting the spectra of aromatic compounds. This makes sense as the differences between sp$^3$- and sp$^2$-hybridized systems are largely in their bonding motifs, which is what CMPNN adds to the description of the molecules.

To demonstrate that our models can predict the spectral differences among conformers, we performed conformer searches using the Merck Molecular Force Field (MMFF94s)[46] in combination with the experimental-torsion knowledge distance geometry method,[47, 48] using RDKit. We generated up to 100 potential conformers per molecule; low-lying conformers with relative energies below 2.5 kcal/mol were selected for further calculations to evaluate their predicted spectra against the corresponding DFT spectra. We selected four representative conformers out of 100 to evaluate their predicted spectra, ensuring a diverse assessment of the model's performance. The comparison between the predicted (CMPNN and DMPNN) and the B3LYP IR molar absorptivity spectra within the 400 − 4000 cm$^{-1}$ range is given in Figure 6 for various conformers of a selected molecule; the comparison for the other three examples is presented in Figures S4 through S9 in the SI.

Figure 6 shows that each conformer has only a slightly different predicted spectrum, especially in the regions where specific vibrational modes are sensitive to the molecular geometry, for instance, around 3000 cm$^{-1}$. The fact that the predictions remain consistent across conformers indicates that the predicted models are robust. This is particularly reassuring in terms of using the predicted molar absorptivity spectra to accelerate the search for transparent hydrocarbon molecules with minimal vibrations in the LWIR energy window ($800 - 1250$ cm$^{-1}$).[3] Thus, the minor variations observed do not undermine the overall performance of the models, which appear to be robust with regard to conformational diversity.

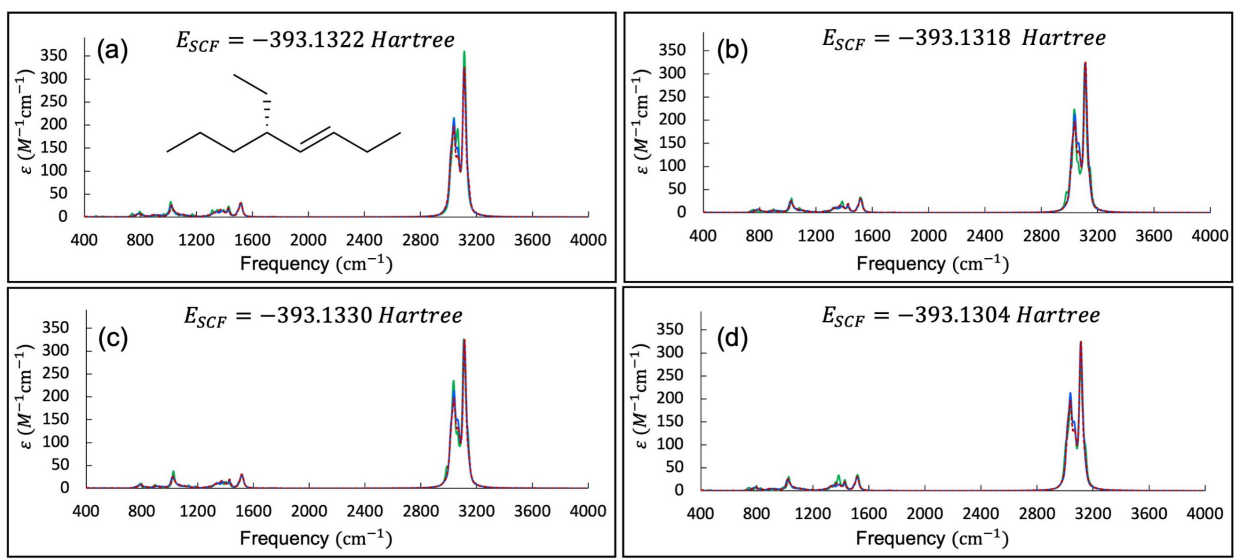

**Figure 6**. Comparison of predicted (CMPNN and DMPNN) versus B3LYP IR molar absorptivity spectra within the $400 - 4000$ cm$^{-1}$ range for various conformers of the selected molecule, whose chemical structure is illustrated in panel (a). The calculated total energies are provided in Hartree for each conformer (since 1 Hartree = 27.21 eV = 627.5 kcal/mol, the total energies of the four conformers are within a range of ~1.6 kcal/mol). An illustration of the four conformations is provided in Figure S3 in the SI.

*(b) Analysis of the model on an external dataset.* To demonstrate the quality of prediction in both CMPNN and DMPNN models, we have evaluated the IR molar absorptivity spectra of about 900 molecules out of the GDB-13 and GDB-11 databases used in the ChemPix[49] package as an external test set. This dataset includes hydrocarbons with more than 10 carbon atoms but excludes molecules with multiple fused rings. Indeed, the external data is incorporated to assess how well the model generalizes to new, unseen data beyond the training set.

The procedure for the DFT calculations and generation of the corresponding spectra follows the workflow shown in Figure 1, with 835 molecules completing this process. Our predictive models were then used to generate the CMPNN and DMPNN IR molar absorptivity spectra for each molecule in the DFT dataset. Figure 7 illustrates the distribution of SIS scores for this external dataset. The CMPNN [DMPNN] model gives an average SIS score of 0.916 [0.911] between the predicted and the reference B3LYP spectra in the $400 - 4000$ cm$^{-1}$ IR window. This result indicates that while there is a slight degradation of the model performance on these previously unseen results for larger molecules, overall, both models are well generalized.

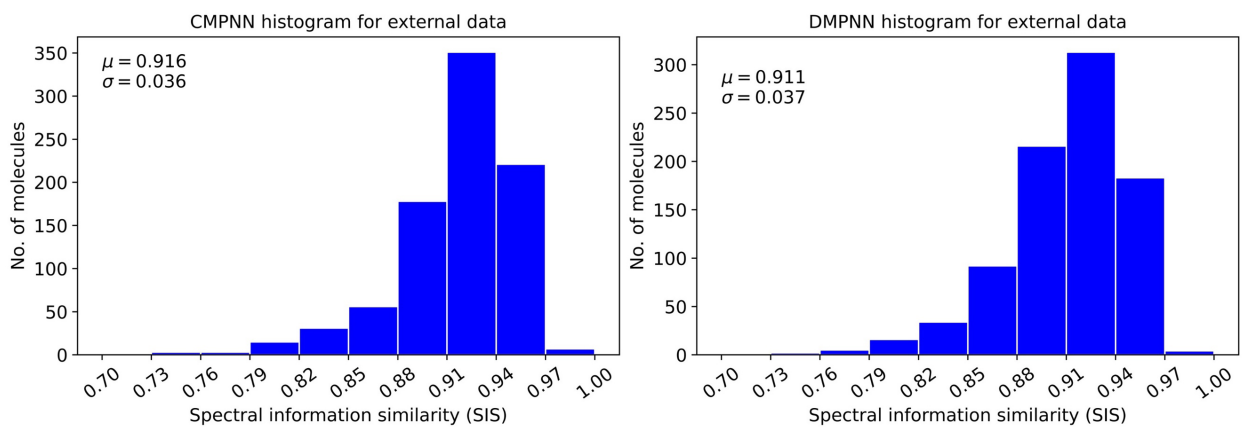

**Figure 7**. Distribution of the SIS scores for the prediction of spectra in the $400 - 4000$ cm$^{-1}$ window in the external dataset; μ and σ represent the average value and standard deviation, respectively.

To address the observed slight degradation in the model's extrapolation capabilities when predicting spectra for hydrocarbon molecules with more than 10 C atoms, we investigate the structural features that may lead to the lower generalizability. Our analysis focuses on the impact of specific structural motifs (SMILES) on the SIS values obtained from the DMPNN model to get a better understanding of how molecular structures affect model performance and predicted spectra.

Table 3 and Table 4 summarize a variety of hydrocarbon substructures and their corresponding impacts on the SIS values. For each substructure, the table details the number of molecules containing the specific motif and the mean SIS values observed within those groups. Additionally, statistical t-tests[50] are included to compare the mean SIS values of two groups (molecules with and without a specific substructure), allowing us to assess whether the presence of each substructure significantly affects the SIS values. The student's t-test[50] determines whether the difference in response between the two groups is statistically significant, with a negative t-statistics indicating that the mean value of the first group is lower than that of the second. The associated p-value represents the probability of obtaining a result at least as extreme as the one shown, if the null hypothesis is true.

Table 3 highlights substructures associated with positive t-statistics, pointing out that these motifs are linked to an increase in SIS values, which suggests better model performance. In contrast, Table 4 focuses on substructures with negative t-statistics, showing that these motifs tend to decrease the SIS values, which is thus related to reduced model performance. What emerges from the results is that the ML model seems to have more difficulties with describing molecules with double bonds, as all motifs that negatively impact the results include them. This further supports the suppositions made for the benzene and naphthalene examples (see Figure 5).

This analysis suggests that, although the ML model tends to predict slightly higher SIS values for the test dataset, it captures similar patterns of variation in both datasets. The statistically significant t-statistics and low p-values across both datasets indicate that the model's predictions are influenced by the same substructural features, reflecting a degree of consistency in its performance.

**Table 3**. Statistical analysis of hydrocarbon substructures and their influence on the SIS values predicted by the DMPNN model. The table includes data for each substructure, such as the number of molecules containing the substructure in both the external and test datasets, mean SIS values, and the results of t-tests (t-statistics and p-value) evaluating the impact of these substructures on SIS values. Positive t-statistics and low p-value (typically less than 0.05) suggest that the presence of a specific substructure is associated with a significant increase in SIS values.

| Num | Substructure | Count (external set) | Mean SIS | T-Statistic | P-Value | Count (test set) | Mean SIS | T-Statistic | P-Value |
|---|---|---|---|---|---|---|---|---|---|
| 1 | 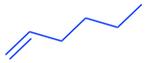 | 599 | 0.91 | 2.49 | 0.01 | 2168 | 0.95 | 5.45 | 0.00 |

| | | | | | | | | |
|---|---|---|---|---|---|---|---|---|
| 2 | 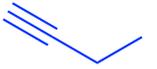 | 414 | 0.91 | 2.34 | 0.02 | 1059 | 0.95 | 3.33 | 0.00 |
| 3 | 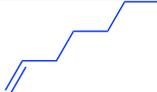 | 418 | 0.92 | 5.45 | 0.00 | 1563 | 0.95 | 4.87 | 0.00 |
| 4 | 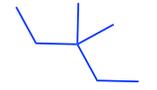 | 194 | 0.92 | 4.91 | 0.00 | 603 | 0.96 | 9.70 | 0.00 |
| 5 | 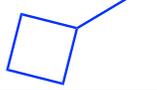 | 93 | 0.93 | 3.75 | 0.00 | 408 | 0.96 | 13.13 | 0.00 |
| 6 | 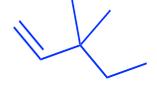 | 143 | 0.92 | 2.51 | 0.01 | 436 | 0.96 | 8.54 | 0.00 |
| 7 | 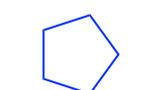 | 56 | 0.93 | 4.77 | 0.00 | 515 | 0.95 | 4.62 | 0.00 |
| 8 | 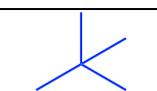 | 282 | 0.92 | 5.49 | 0.00 | 841 | 0.96 | 9.87 | 0.00 |
| 9 | 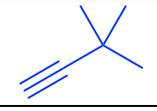 | 62 | 0.92 | 3.69 | 0.00 | 136 | 0.95 | 2.95 | 0.00 |
| 10 | 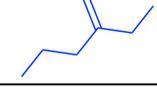 | 260 | 0.92 | 2.51 | 0.01 | 795 | 0.95 | 5.70 | 0.00 |

**Table 4**. Analysis of hydrocarbon substructures and their impact on the SIS values predicted by the DMPNN model. This table presents a statistical analysis of the selected substructures and their influence on the SIS scores. For each substructure, the table includes the count of molecules containing the substructure, the mean SIS values, and the t-statistics with corresponding p-value from t-tests assessing the impact of the substructure on the SIS values. Negative t-statistics and low p-value (typically less than 0.05) suggest that the presence of a specific substructure is associated with a significant decrease in SIS values.

| Num | Substructure | Count (external set) | Mean SIS | T-Statistic | P-Value | Count (test set) | Mean SIS | T-Statistic | P-Value |
|---|---|---|---|---|---|---|---|---|---|
| 1 | 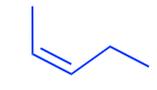 | 647 | 0.91 | -4.66 | 0.00 | 2232 | 0.94 | -4.18 | 0.00 |

| 2 | 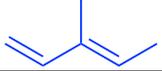 | 269 | 0.89 | -6.6 | 0.00 | 647 | 0.94 | -3.17 | 0.00 |
|---|---|---|---|---|---|---|---|---|---|
| 3 | 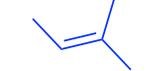 | 495 | 0.90 | -2.84 | 0.00 | 1354 | 0.95 | 4.31 | 0.00 |
| 4 | 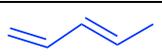 | 370 | 0.90 | -8.60 | 0.00 | 1018 | 0.93 | -9.48 | 0.00 |
| 5 | 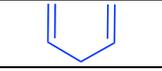 | 337 | 0.90 | -6.91 | 0.00 | 967 | 0.94 | -7.09 | 0.00 |
| 6 | 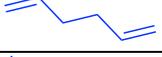 | 272 | 0.90 | -2.61 | 0.00 | 721 | 0.94 | -4.90 | 0.00 |
| 7 | 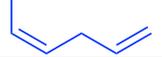 | 290 | 0.89 | -6.5 | 0.00 | 792 | 0.94 | -6.34 | 0.00 |
| 8 | 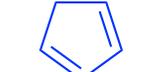 | 77 | 0.90 | -2.49 | 0.01 | 220 | 0.94 | -2.83 | 0.01 |
| 9 | 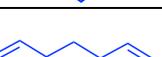 | 164 | 0.90 | -2.52 | 0.01 | 519 | 0.94 | -5.86 | 0.00 |
| 10 | 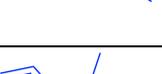 | 57 | 088 | -3.80 | 0.00 | 92 | 0.93 | -3.49 | 0.00 |

*(c) Analysis of the model using a principal component analysis (PCA).* Principal component analysis (PCA)[51] is a well-established unsupervised dimensionality reduction approach, applied successfully for instance in catalysis design[52, 53] and drug design,[54] which we exploit here to identify the most significant structural features in the prediction of spectra. Applying PCA onto our datasets reveals that we can reduce the atom-only feature set to three principal components and the atom + bond features to four principal components. Figure 8 shows the variance explained by each principal component in the PCA approach. The summary information indicates that the first, second, and third principal components of the atom features (PCa1, PCa2, and PCa3) explain 69.11%, 19.71%, and 6.06% (total of 94.88%) of the variance, respectively. The variances explained by the first four principal components of the bond features (PCab1, PCab2, PCab3, and PCab4) are 48.64%, 25.78%, 7.46%, and 5.47%, respectively; the cumulative explained variance for these four components is 87.35%, indicating that they account indeed for most of the variance in atom + bond features. Using the first two principal components also confirms that the current descriptor space well characterizes the relevant molecular space as the external dataset is within the convex hull of the test set, as shown in Supplementary Figure S10. Overall, these results suggest that the atom-only features provide a more compact description of the variance in the system, with both representations being adequate to describe the systems.

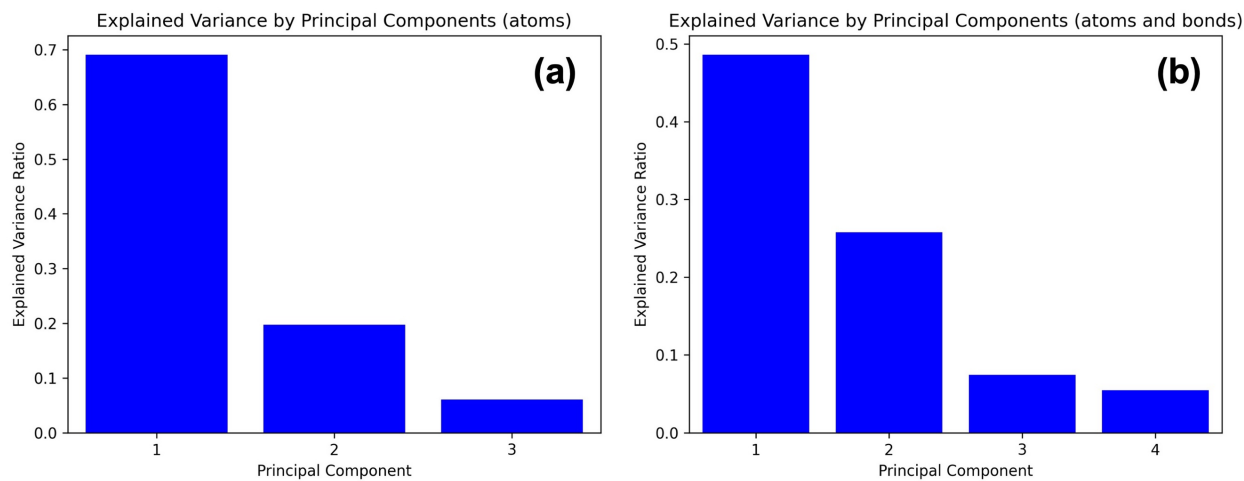

**Figure 8**. Illustration of the variance explained by principal components in a principal component analysis (PCA). Each bar shows the percentage of total variance in the dataset accounted for by an individual principal component, ordered by their significance. Panel (a) shows the variance explained by principal components of atom features while (b) shows the variance explained by principal components of combined atom and bond features.

To identify the key features in the prediction of the IR spectra, we examined the loadings of the features on the components. Here, we consider an absolute loading value over 0.1 as an important contributor. For all three atom-based principal component loadings, features that represent the number of bonds linked to each atom and hybridization are critical, with the atomic number being also important for PCa1. For the bond features, a subset of features that describe whether the bond is in a ring, the type of bond (*i.e.*, single, double, or triple), and if the bond is conjugated are the important loadings in all principal components. The PCA analysis indicates that stereochemical features (information such as a cis/trans double bond) play no role in the improvement of the model. The atom and bond loadings are presented in Figure S2 of the SI, with the relevant values available therein.

Since PCA provides insight into which combination of original features (principal components) are important for prediction, we trained a model using DMPNN and these reduced atom and bond features while using the same training procedure. Results in Figure 9 suggest that the reduced features coming out of the PCA analysis effectively capture the most relevant information while reducing dimensionality.

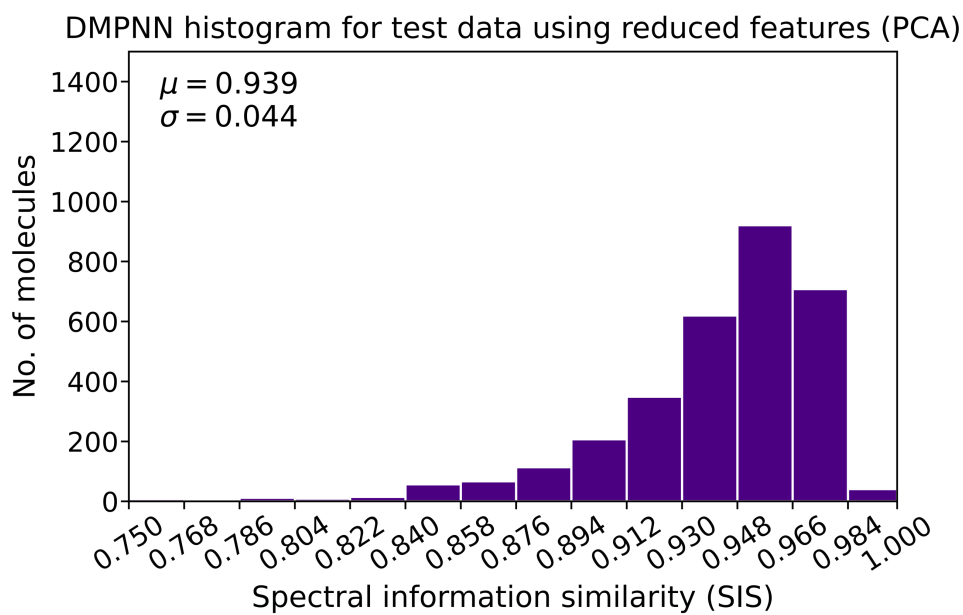

**Figure 9**. Distribution of the SIS scores for the IR spectra prediction in the test split using the reduced features coming from the principal component analysis (PCA) and DMPNN architectures; μ and σ represent the average value and standard deviation, respectively.

Considering the atom and bond loadings obtained by the PCA analysis, we used the features given in Table 5 and Table 6 to train our models. Figure 10 presents the distribution of the SIS scores for the IR spectra prediction in the test split. An average SIS value of 0.943 [0.942] and a standard deviation of 0.044 [0.045] in the CMPNN [DMPNN] procedure indicate that the accuracy of the predicted models is preserved while considering only important features. Our study demonstrates that both algorithms achieve comparable predictive performance in the case of hydrocarbons and suggests that either model could be a viable option in future applications.

**Table 5**. Important atom features in the context of our models.

| Feature | Description | size |
| --- | --- | --- |
| Atom type | Type of atom (ex. C, N, O), by atomic number | 100 |
| # bonds | Number of bonds the atom is involved in | 6 |
| Hybridization | sp, $sp^2$, $sp^3$, $sp^3d$, or $sp^3d^2$ | 5 |
| Atomic mass | Mass of the atom, divided by 100 | 1 |

**Table 6.** Important bond features in the context of our models.

| Feature | Description | size |
| --- | --- | --- |
| Bond type | Single, double, triple, or aromatic | 4 |
| Conjugated | Whether the bond is conjugated | 1 |
| In ring | Whether the bond is part of a ring | 1 |

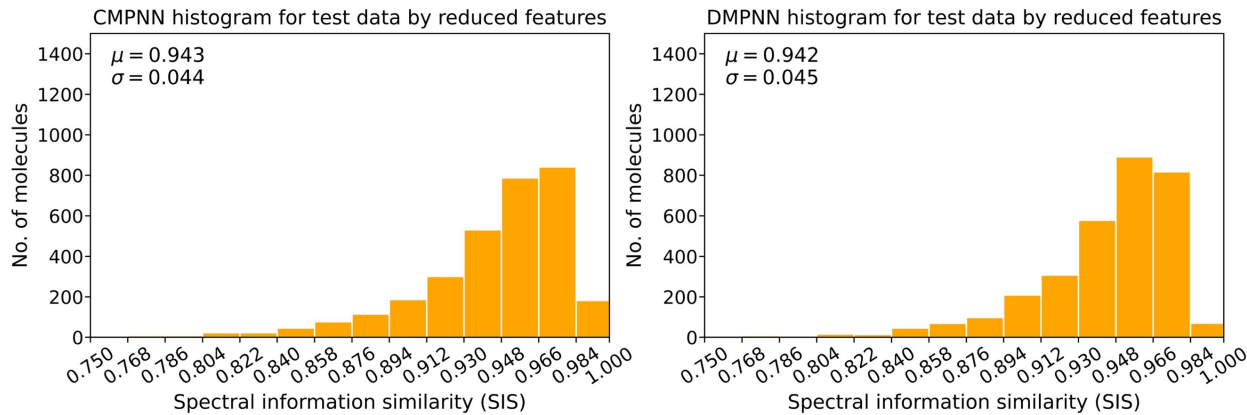

**Figure 10**. Distribution of the SIS scores for the IR spectra prediction in the test split using reduced features along with CMPNN and DMPNN architectures; μ and σ represent the average value and standard deviation, respectively.

## Application of predicted molar absorptivity spectra for accelerated discovery of hydrocarbon molecules in IR materials design

An important application of our predictive models is their ability to identify hydrocarbon molecules with minimal / weak absorption across different IR regions within a vast chemical space. For instance, the design of LWIR-transparent organic materials is challenging due to the fact that nearly all organic molecules absorb in this spectral window, known as the 'IR Fingerprint Region.' Consequently, fundamental research into the design of LWIR-transmissive polymeric materials is essential for developing a new class of cost-effective LWIR plastic optics,[55] which can be used to create more affordable IR imaging systems. In our previous work,[3] we integrated DFT calculations and high-throughput screening to simulate molar absorptivity IR spectra and design novel hydrocarbon molecules with enhanced LWIR optical transparency that can be exploited for producing novel transparent hybrid sulfur/organic plastic materials. A message from our present work is that the use of predictive models with machine learning (ML) can significantly accelerate this search process by efficiently scanning large chemical spaces for suitable comonomers with minimal LWIR absorption. Our approach offers a major scientific advantage over traditional methods by accurately predicting molar absorptivity spectra, rather than just normalized spectra. This is crucial for calculating transparency in the LWIR region, making it a valuable tool for advancing the development of LWIR transparent materials.

**Conclusion**

We reported a predictive model based on a directed graph-based communicative message passing neural network (CMPNN) architecture to predict the IR absorptivity including absolute intensities (and not only relative intensities) using a dataset of 31,570 hydrocarbons considering ten or fewer carbon atoms. The training was performed on the basis of the learned molecular representations in the open-source Chemprop-IR software. The loss function of spectral information divergence (SID) was modified in view of the prediction of IR absorbance intensities. We also considered a directed graph-based message passing neural network (DMPNN) baseline model trained by the same dataset. The predicted spectra based on both models were compared to the IR spectra calculated at the first-principles DFT/B3LYP level.

Very similar performance is achieved by both architectures, which underlines that strengthening the message interactions between edges and nodes in CMPNN does not necessarily improve the prediction of the spectra in graph-based message passing neural networks. Importantly, both predictive models offer high-fidelity performance in reproducing the DFT B3LYP spectra in the test set. To examine the transferability of the models could be extended to larger molecules, we considered an external dataset of about 900 hydrocarbon molecules containing up to 13 carbon atoms. The outcome indicates that the ML models are sensitive to the size difference between the training set and the test molecules. Overall, we emphasize that our ML model is capable of predicting IR absorbances with the accuracy of first-principles calculations, which provides the prospect to accelerate the search of new high-performing hydrocarbons, for instance, in the content of imaging applications in the mid-wave and/or long-wave IR domains.

We applied a principal component analysis to identify the features that are most significant for predicting spectra. By training a model with PCA-reduced features using the DMPNN architecture ,we demonstrated that these condensed features are sufficient to achieve an accuracy comparable to that obtained when using the complete feature set in Chemprop-IR.

Next, we used the significant atom and bond features identified by PCA to train models with both CMPNN and DMPNN architectures. Our findings suggest that these selected features are crucial in capturing the key vibrational properties of hydrocarbon materials in this study.

Since CMPNN strengthens interactions between edges (bonds) and nodes (atoms) vs. DMPNN, the only slight improvement or equivalent performance that we find for CMPNN vs. DMPNN implies that the bond feature interactions are already sufficiently captured in CMPNN or that the additional interactions in CMPNN provide only marginal benefits over DMPNN. Thus, given that overall CMPNN and DMPNN perform similarly in predicting IR spectra, either approach could be effectively exploited in subsequent studies and applications of these methodologies on hydrocarbon systems.

**Timing**

We used 28 cores of an AMD EPYC 7642 (Rome) and an Nvidia Volta V100 GPU for training. Training an ensemble of 10 sub-models takes roughly 48 hours. Spectral prediction for 3,366 molecules in the test split using both models takes about 1 minute from the ensemble of 10 sub-models.

## Associated content

## Data availability statement

All data needed to evaluate the conclusions of this work are presented within the paper and/or the Supporting Information. Additional data, including the source code associated with the CMPNN approach based on the Chemprop-IR software, trained models using the CMPNN and DMPNN protocols, training, validation, and test datasets, as well as the external dataset and SIS values for this study, are all available on Zenodo[56] [10.5281/zenodo.13844304]. The list of SMILES strings and the optimized XYZ coordinates used to simulate the DFT molar absorptivity IR spectra are available on Zenodo[57] [10.5281/zenodo.13163525].

## Acknowledgments

This work was funded by the National Science Foundation and the Air Force Research Laboratory through award No. DMREF-2118578, as well as by the College of Science of The University of Arizona. The authors are thankful for the use of the High Performance Computing (HPC) resources supported by the University of Arizona TRIF, UITS, and Research, Innovation, and Impact (RII) Offices and maintained by the Research Technologies Department.

# SUPPLEMENTARY INFORMATION

## Prediction of the Infrared Absorbance Intensities and Frequencies of Hydrocarbons: A Message Passing Neural Network Approach

**Maliheh Shaban Tameh, Veaceslav Coropceanu,**

**Thomas A.R. Purcell[*], and Jean-Luc Brédas[*]**

Department of Chemistry and Biochemistry

The University of Arizona

Tucson, Arizona 85721-0041

* Emails: purcellt@arizona.edu; jlbredas@arizona.edu

*Communicative message passing neural network (CMPNN).* In this section, an overview of the molecular embedding generation with CMPNN is described.[1] A comparison between generic MPNN,[2] Directed MPNN (DMPNN)[3], and CMPNN is listed in Table S1, which indicates that MPNN uses node-hidden states $h(v)$ and messages $m(v)$, DMPNN operates on edge-hidden states $h(e_{v,w})$ and messages $m(e_{v,w})$, while CMPNN is based on $h(v)$, $m(v)$, $h(e_{v,w})$, and $m(e_{v,w})$.

**Table S1**. Differences between the messages and the hidden states sent in the message passing phase of generic MPNN, Directed MPNN (DMPNN), and CMPNN algorithms.

| Input | MPNN | DMPNN | CMPNN |
|---|---|---|---|
| Node-hidden states | $h(v)$ | | $h(v)$ |
| Message associated with nodes | $m(v)$ | | $m(v)$ |
| Edge-hidden states | | $h(e_{v,w})$ | $h(e_{v,w})$ |
| Message associated with edges | | $m(e_{v,w})$ | $m(e_{v,w})$ |

A CMPNN algorithm operates on a directed molecular graph $G = (V, E)$ with node (atom) features $x_v$ ($\forall\ v \in V$) and edge (bond) features $x_{e_{v,w}}$ ($\forall\ e_{v,w} \in E$). The algorithm includes two phases: a message passing phase and a readout phase. In the former, the node and edge information of the molecule is propagated at each iteration for T time steps to build a neural representation of the molecule. The directed message passing procedure in CMPNN is sketched in Figure S1. In the readout phase, a readout function is used to make a prediction based on the final hidden states by computing a feature vector for the whole graph. All the functions used in the process are differentiable.

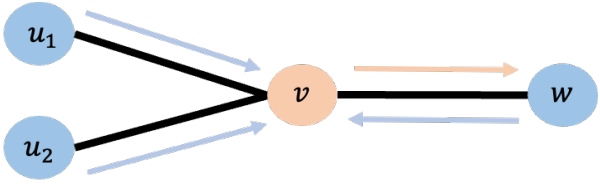

**Figure S1**. Directed message passing.[1]

In the following, we summarize the algorithm as was described by Song et al.[1]

$h^0(v) \leftarrow x_v, \forall\, v \in V; h^0(e_{v,w}) \leftarrow x_{e_{v,w}}, \forall\, e_{v,w} \in E.$

For $t = 1, \ldots, T$ do
    For $v \in V$ do

$$m^{t+1}(v) = \sum_{u \in N(v)} h^t(e_{u,v}) \quad (1)$$

$$h^{t+1}(v) = h^t(v).m^{t+1}(v) \quad (2)$$

    End for

    For $e \in E$ do

$$m^{t+1}(e_{v,w}) \leftarrow h^{t+1}(v) - h^t(e_{w,v}) \quad (3)$$

$$h^{t+1}(e_{v,w}) = U_t\left(h^t(e_{v,w}), m^{t+1}(e_{v,w})\right) = U\left(h^t(e_{v,w}), m^{t+1}(e_{v,w})\right) = \sigma(h^0(e_{v,w}) + W.m^{t+1}(e_{v,w})) = \sigma(h^0(e_{v,w}) + W_m m^{t+1}(e_{v,w})) \quad (4)$$

    End for
End for

Equation 3 in D-MPNN is given as $m^{t+1}(e_{v,w}) \leftarrow h^t(e_{u,v}), \forall\, u \in N(v)\backslash w$. In equation (4), $U_t$ is the edge update function and we implement $U_t$ using the same neural network on every step. $W_m \in \mathbb{R}^{h \times h}$ is a learned matrix with hidden size $h$. By adding $h^0(e_{v,w})$ at each step, we provide a skip connection to the original feature vector for that edge. $\sigma$ is the rectified linear unit (ReLU) activation function.[4]

$$\boldsymbol{m}(v) = \sum_{u \in N(v)} \boldsymbol{h}^T(\boldsymbol{e}_{u,v}) \tag{5}$$

$$\boldsymbol{h}(v) \leftarrow \text{COMMUNICATE}\,(\boldsymbol{m}(v), \boldsymbol{h}^T(v), \boldsymbol{x}(v)) \tag{6}$$

Equation 6 is the final atom representation of the molecule given by a communicative function. It indicates the messages from incoming bonds, the current atom's representation, and the atom's initial information. In the context of node-edge message communication modules, we use inner product kernel as described by Song et al.[1]

$$\boldsymbol{z} \leftarrow \text{Readout}\,(\boldsymbol{h}(v), \forall\, v \in V) \tag{7}$$

Equation 7 indicates the readout phase and is given here as

$$z = \Sigma_{v \in V} GRU(h), \tag{8}$$

where $h = \Sigma_{v \in G} h(v)$ is the sum of the atom hidden states to obtain a feature vector for the molecule. GRU represents the Gated Recurrent Unit introduced by Cho et al.[5]

At the end, the property predictions are generated as $\hat{y} = f(h)$, where $f(.)$ is a feed-forward neural network.

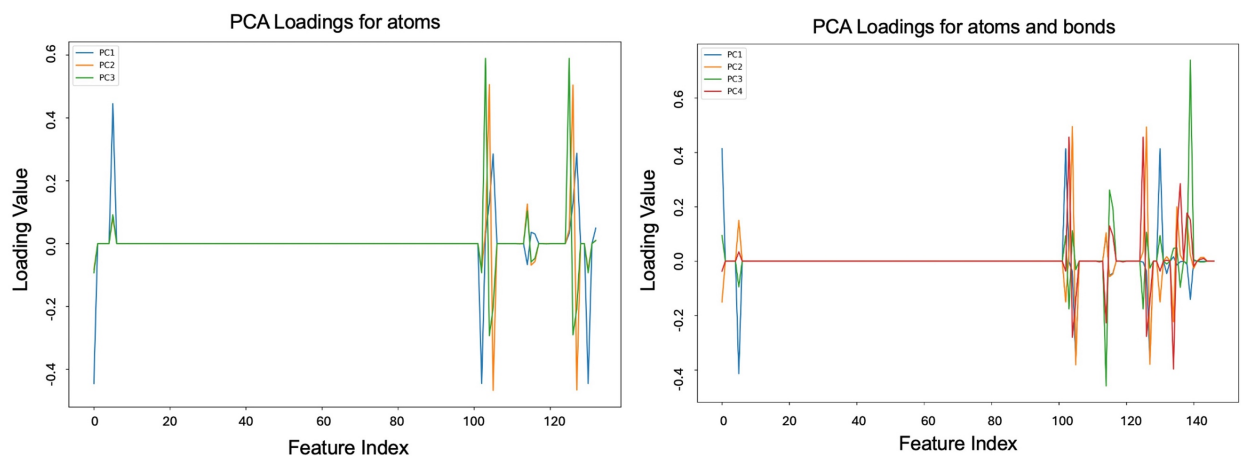

**Figure S2**. Comparison of PCA loadings for atom features and combined atom-bond features. The left plot displays the PCA loadings for atom features only, where the three principal components (PCa1, PCa2, and PCa3) show the contribution of each atomic feature across the feature index. The right plot includes both atom and bond features, showing the loadings for the first four principal components (PCab1, PCab2, PCab3, and PCab4). The feature indices correspond to various atomic and bond attributes. The loading values highlight which features contribute most to the variance captured by each principal component.

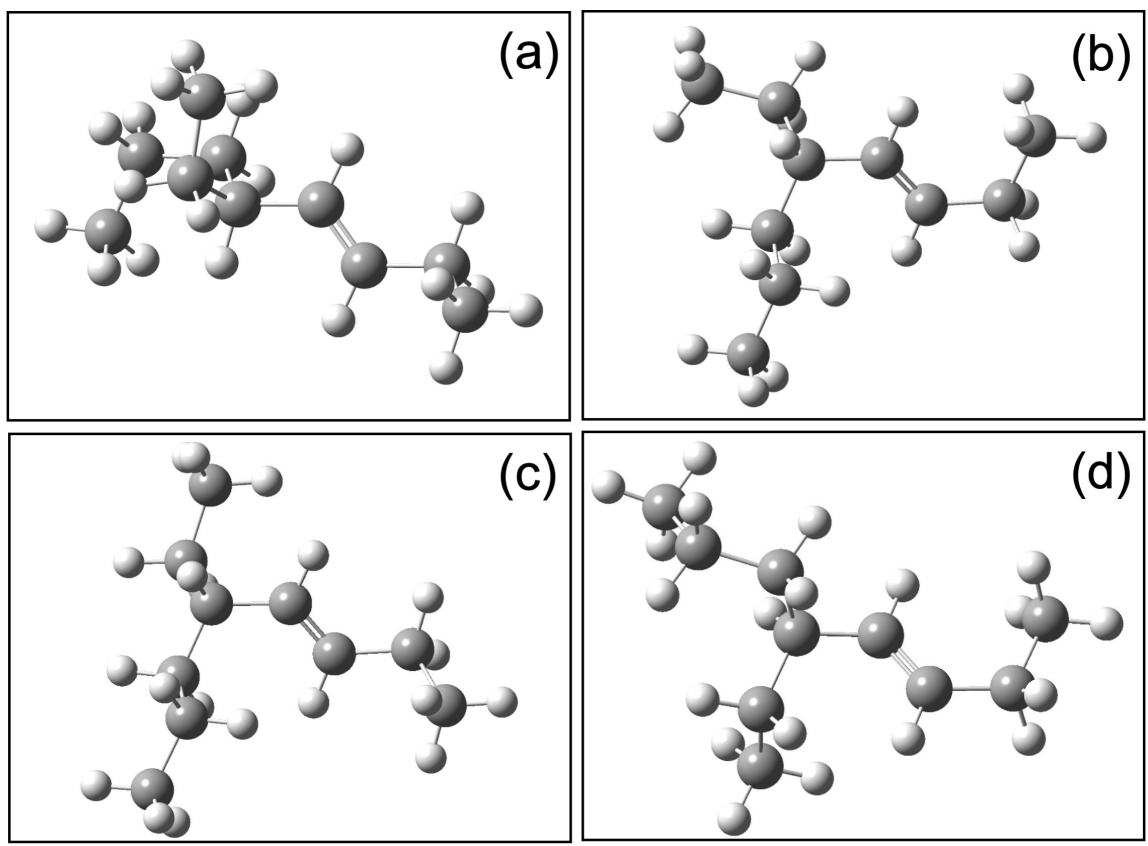

**Figure S3.** Illustration of the four conformations presented in Figure 6 of the main paper. Panels (a), (b), (c), and (d) correspond to the respective conformations (a), (b), (c), and (d) in Figure 6.

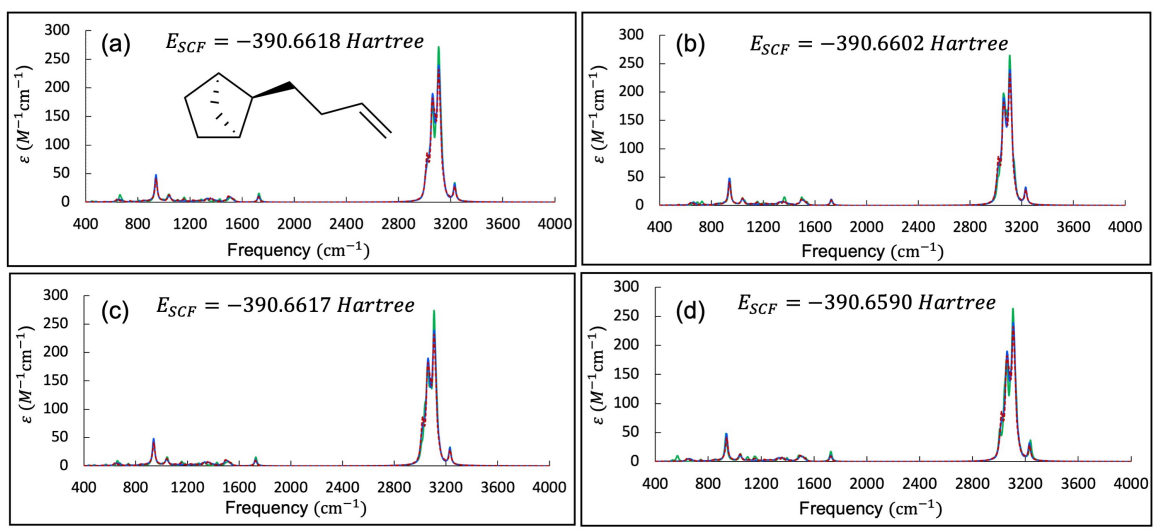

**Figure S4**. Comparison of predicted (CMPNN and DMPNN) versus B3LYP IR molar absorptivity spectra within the $400 - 4000 \text{ cm}^{-1}$ range for various conformers of the selected molecule, whose chemical structure is illustrated in panel (a). The calculated total energies are provided in Hartree for each conformer (since 1 Hartree = 27.21 eV = 627.5 kcal/mol, the total energies of the four conformers are within a range of ~1.7 kcal/mol). An illustration of the four conformations is provided in Figure S5.

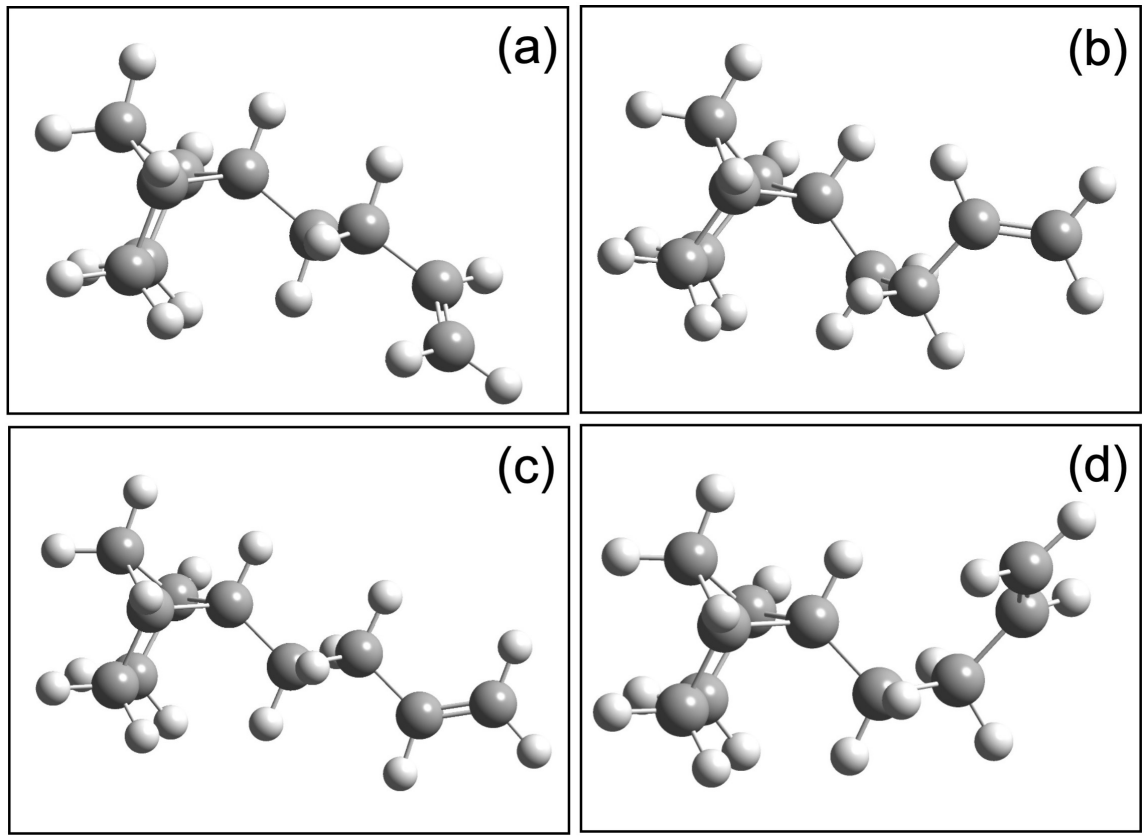

**Figure S5.** Illustration of the four conformations presented in Figure S4. Panels (a), (b), (c), and (d) correspond to the respective conformations (a), (b), (c), and (d) in Figure S4.

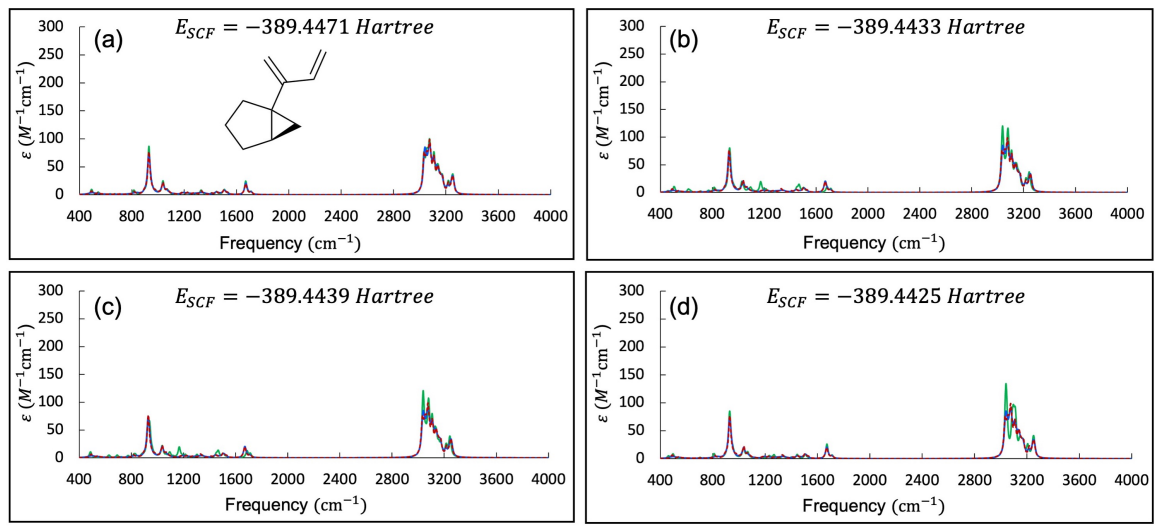

**Figure S6**. Comparison of predicted (CMPNN and DMPNN) versus B3LYP IR molar absorptivity spectra within the $400 - 4000$ cm$^{-1}$ range for various conformers of the selected molecule, whose chemical structure is illustrated in panel (a). The calculated total energies are provided in Hartree for each conformer (since 1 Hartree = 27.21 eV = 627.5 kcal/mol, the total energies of the four conformers are within a range of ~2.9 kcal/mol). An illustration of the four conformations is provided in Figure S7 in the SI.

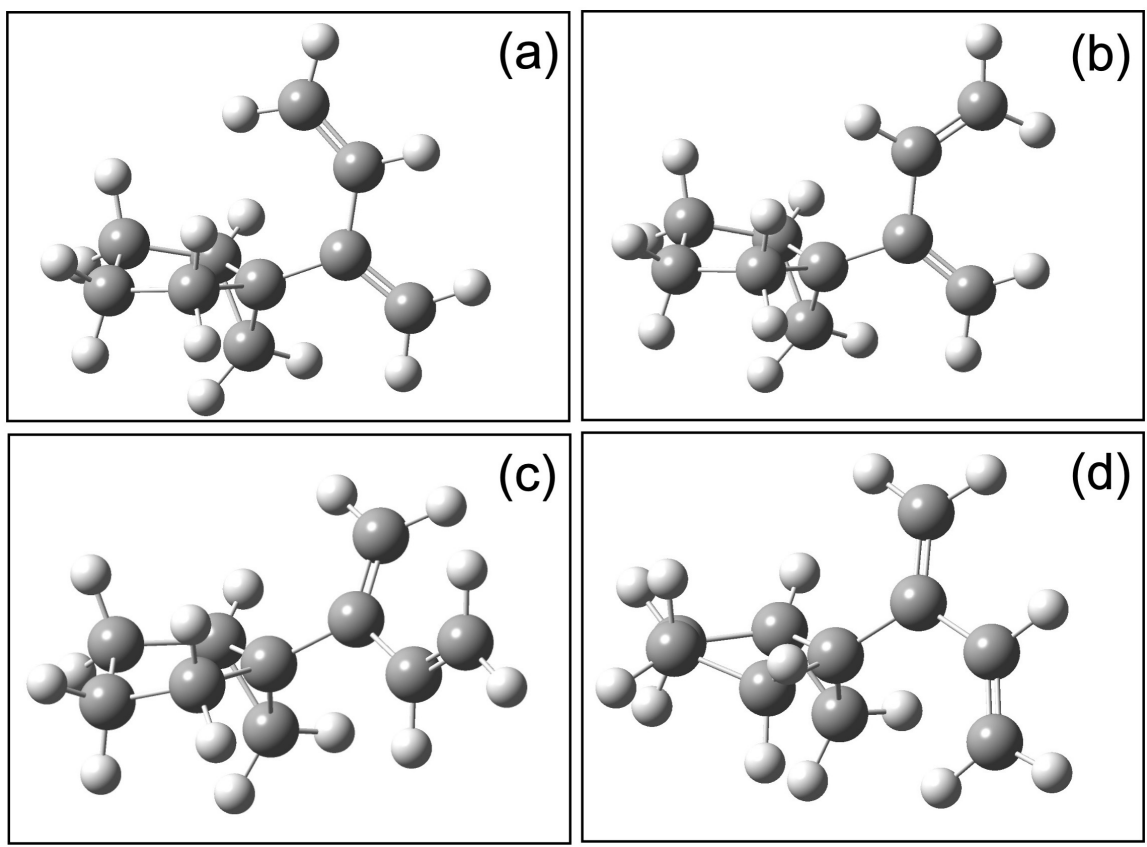

**Figure S7.** Illustration of the four conformations presented in Figure S6. Panels (a), (b), (c), and (d) correspond to the respective conformations (a), (b), (c), and (d) in Figure S6.

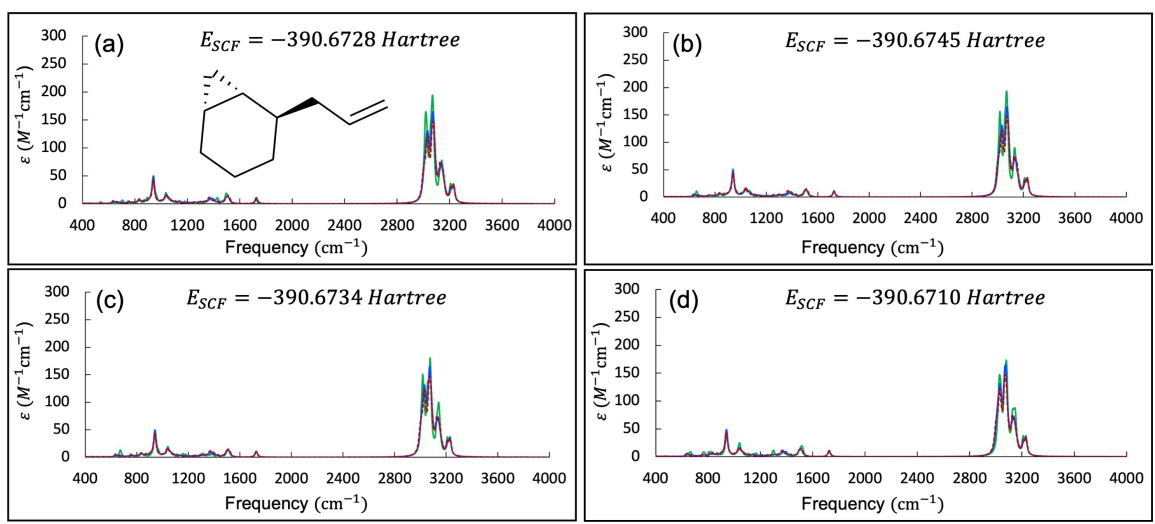

**Figure S8**. Comparison of predicted (CMPNN and DMPNN) versus B3LYP IR molar absorptivity spectra within the $400 - 4000$ cm$^{-1}$ range for various conformers of the selected molecule, whose chemical structure is illustrated in panel (a). The calculated total energies are provided in Hartree for each conformer (since 1 Hartree = 27.21 eV = 627.5 kcal/mol, the total energies of the four conformers are within a range of ~2.2 kcal/mol). An illustration of the four conformations is provided in Figure S9.

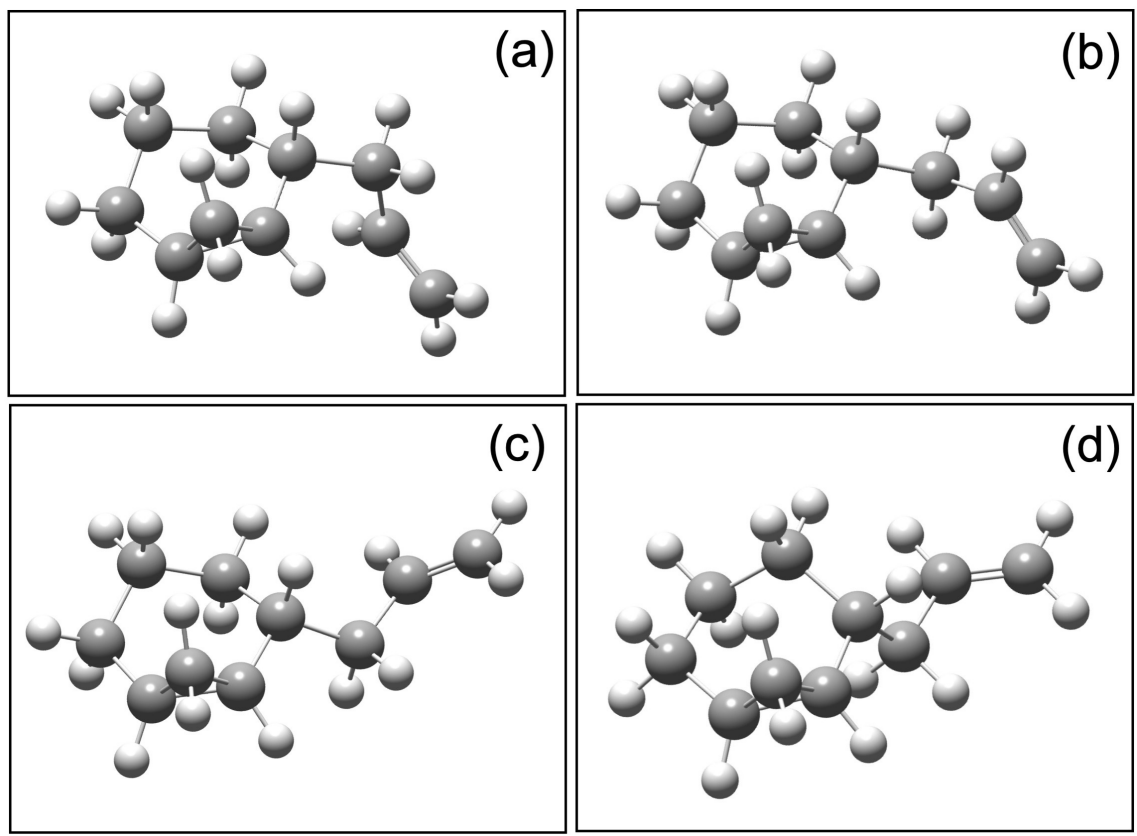

**Figure S9.** Illustration of the four conformations presented in Figure S8. Panels (a), (b), (c), and (d) correspond to the respective conformations (a), (b), (c), and (d) in Figure S8.

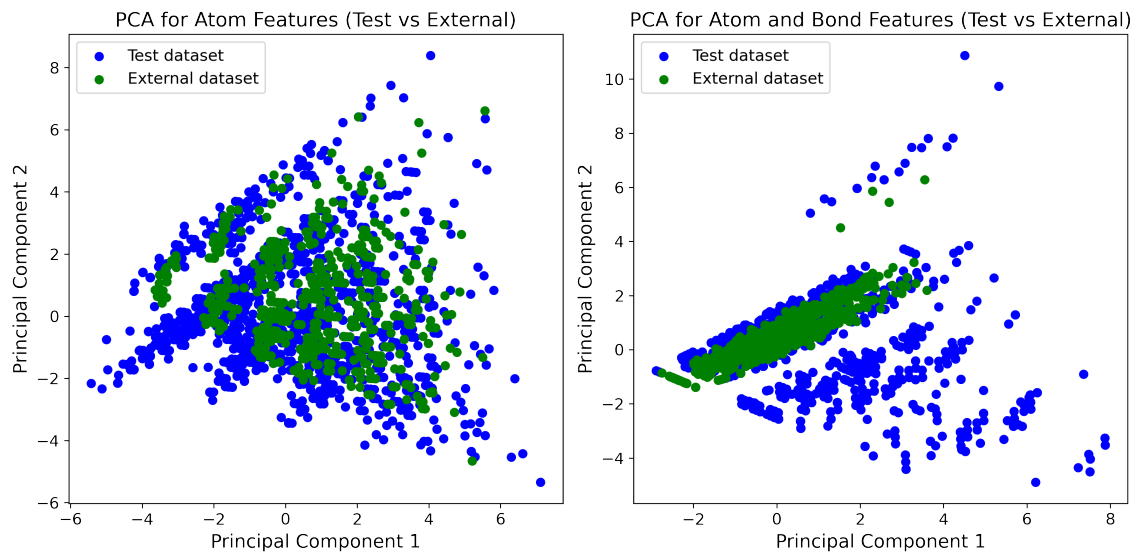

**Figure S10.** PCA scatter plots comparing atom and atom + bond features between the test dataset and external dataset. The left plot shows an apparent overlap between the test dataset (blue) and the external dataset (green), with both occupying similar regions in PCA space. In the right plot, the external dataset is confined to a more specific region, while the test dataset is spread out more widely, suggesting that the external dataset introduces new atomic and bonding environments that are not fully represented in the test dataset.